\documentclass[superscriptaddress,nofootinbib,amsmath,amssymb,aps,prx,twocolumn]{revtex4-2}
\usepackage{graphicx}
\usepackage{dcolumn}
\usepackage{bm}
\usepackage{latexsym}
\usepackage{amsmath, dsfont, physics}
\usepackage[colorlinks=true, linkcolor=blue, citecolor=blue, urlcolor=blue]{hyperref}
\usepackage{amssymb}
\usepackage{graphicx}
\usepackage{caption}
\usepackage{subfigure}
\usepackage{float}
\usepackage{mathrsfs}
\usepackage{color}
\usepackage{txfonts}
\usepackage[justification=centering,
            format=plain]{caption}
\renewcommand{\raggedright}{\leftskip=0pt \rightskip=0pt plus 0cm}

\begin{document}

\title{Engineering long-lived entanglement through dissipation in quantum hybrid solid-state platforms}

\author{Jayakrishnan M. P. Nair}
\email{muttathi@bc.edu}

\affiliation{Department of Physics, Boston College, 140 Commonwealth Avenue, Chestnut Hill, Massachusetts 02467, USA}

\author{Benedetta Flebus}
\email{flebus@bc.edu}
\affiliation{Department of Physics, Boston College, 140 Commonwealth Avenue, Chestnut Hill, Massachusetts 02467, USA}

\date{\today}

\begin{abstract}
Spin squeezing, a form of many-body entanglement, is a crucial resource in quantum metrology and information processing. While experimentally viable protocols for generating stable spin squeezing have been proposed in quantum optics setups, there is growing interest in quantum hybrid solid-state systems as alternative platforms for both engineering and exploring many-body quantum phenomena. In this work, we propose a scheme to generate long-lived spin squeezing in an ensemble of solid-state qubits interacting with electromagnetic noise emitted by a squeezed solid-state bath. We identify the conditions under which quantum correlations within the bath can be transferred to the qubit array, driving it into an entangled state independently of its initial configuration. To assess the experimental feasibility of our approach, we analyze the dynamics of an array of solid-state spin defects coupled to a common ferromagnetic bath, which is driven into a non-equilibrium squeezed state through its interaction with a surface acoustic wave mode. Our results demonstrate that the ensemble can exhibit steady-state spin squeezing under suitable conditions,  opening new pathways for the generation of robust many-body entanglement in solid-state spin ensembles.
\end{abstract}

\maketitle

\section{Introduction}

Central to the development of quantum technologies is the concept of squeezing, which characterizes the reduction of quantum uncertainty in one observable of a system at the expense of increased uncertainty in its conjugate observable. The generation of squeezed states has been explored in a broad variety of systems, including — but not limited to — Bose-Einstein condensates \cite{orzel2001squeezed,esteve2008squeezing,gross2010nonlinear,riedel2010atom,sorensen2001many,law2001coherent,poulsen2001positive,raghavan2001generation,jenkins2002spin,jaksch2002dynamically,jing2002mutual,sorensen2002bogoliubov,micheli2003many,molmer2003quantum}, optomechanical systems \cite{PhysRevA.91.013834,wollman2015quantum,nunnenkamp2010cooling,purdy2013strong,agarwal2016strong}, mechanical systems \cite{marti2024quantum, PhysRevLett.107.213603}, and magnonic systems \cite{kamra2020magnon,li2019squeezed,PhysRevLett.116.146601,PhysRevB.96.020411,PhysRevB.100.174407,PhysRevB.101.014416}. Among these squeezing realizations, a valuable resource for quantum metrology lies in spin squeezed states~\cite{ma2011quantum}, which can be realized in systems that can be described by collective spin variables, such as the atomic ensembles \cite{kuzmich1997spin,appel2009mesoscopic,louchet2010entanglement,takano2009spin} conventionally used in Ramsey spectroscopy \cite{wineland1992spin,wineland1994squeezed} and atomic clocks \cite{wineland1994squeezed,bigelow2001squeezing,appel2009mesoscopic}.

An essential property of spin squeezing is its direct relation with multi-partite entanglement, whose characterization and measurement remain otherwise elusive in many-body systems. The degree of spin squeezing can be quantified, among others \cite{PhysRevA.47.5138}, in terms of Wineland's spin squeezing parameter $\xi_R^2$ \cite{wineland1994squeezed}, which compares the quantum fluctuations in one spin component to those of a coherent spin state. Specifically, a value of $\xi_R^2$ less than one indicates reduced quantum fluctuations and, for any many-body spin-1/2 state, signals the presence of entanglement \cite{sorensen2001many}. 

Several proposals for generating spin squeezing in atomic ensembles rely on their interaction with a common squeezed photonic reservoir, which enables the transfer of squeezing from the light to the atoms. For instance, in the far-detuned regime, the coherent interactions between the atomic ensemble and a cavity mode result in an effective qubit-qubit Hamiltonian whose nonlinear twisting terms can be used to generate spin squeezing. Such one- and two-axes twisting Hamiltonians have been explored both theoretically \cite{PhysRevLett.116.053601, PhysRevLett.127.210401} and experimentally \cite{hosten2016quantum,leroux2010implementation} across various platforms. As for any entanglement protocol relying on coherent qubit-qubit interactions, however, here the decoherence resulting from the dissipative interaction with the environment plays a detrimental role, shortening the lifetime of spin squeezed states. 

\begin{figure*}
\captionsetup{justification=raggedright,singlelinecheck=false}
\centering
   \includegraphics[scale=1.11]{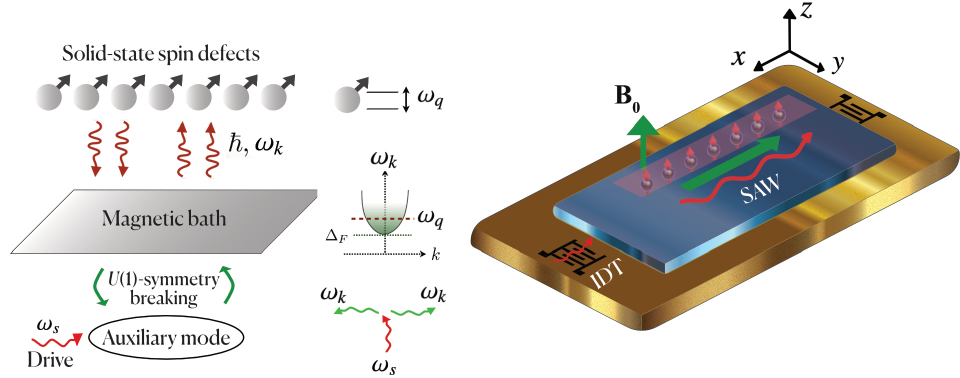}
\caption{Left: Schematic illustration of the proposed scheme for generating steady-state spin squeezing in an ensemble of solid-state spin defects interacting with a magnetic bath. A nonlinear, $U(1)$-symmetry breaking, interaction between the magnetic reservoir and an auxiliary mode, classically driven at frequency $\omega_{s}$, generates non-equilibrium squeezed states, i.e., correlated magnon pairs with frequency $\omega_{k} \simeq \omega_s/2$, within the bath. When the resonance frequency $\omega_{q}$ of the solid-state spin defects is tuned above the spin-wave gap $\Delta_{F}$, i.e., $\omega_{q}>\Delta_{F}$, the ensemble can simultaneously absorb and emit correlated magnon pairs from and into the bath, which, in turn, facilitates the buildup of steady-state spin squeezing within the ensemble. Right: A possible realization of our scheme. An array of solid-state spin defects is set above a magnetic sample placed on top of a piezoelectric crystal. Two interdigital transducers (IDTs) generate SAWs that propagate through the crystal, periodically applying strain to the magnet. A static magnetic field $\mathbf{B}_0$ aligns the magnetic order in the ferromagnetic bath along the $\hat{z}$ direction.}
\label{sch}
\end{figure*}
An alternative route is offered by reservoir engineering approaches that turn dissipation into a resource for the generation and stabilization of a form of many-body quantum entanglement that is robust to variations in the system's initial state \cite{PhysRevLett.110.120402,PhysRevX.12.011015,PhysRevResearch.4.013089,PhysRevLett.70.556,PhysRevA.88.063833,PhysRevA.89.013820,wollman2015quantum,kienzler2015quantum,PhysRevX.5.041037,PhysRevLett.115.243601,PhysRevLett.117.100801,PRXQuantum.2.020323}. In these schemes, the squeezing properties of the bath are imparted to the system via its dissipative coupling to a nontrivial common environment, with an efficiency that increases with the size of the qubit ensemble. The dissipative preparation of many-body spin squeezed states has been proposed in several settings, e.g., involving direct driving with squeezed light \cite{agarwal1989nonequilibrium, PhysRevA.41.3782,PhysRevA.49.4968}, Raman processes in structured atoms~ \cite{PhysRevLett.110.120402}, and coupling to  bosonic modes that, in turn, interact with a squeezed reservoir \cite{PhysRevX.12.011015}. However, it still awaits experimental realization. 

The central goal of this work is to transfer and adapt this approach to solid-state quantum hybrid platforms, consisting of an ensemble of solid-state spin defects interacting with a common magnetic bath, which, as we will demonstrate, can be driven into a squeezed state using already established experimental techniques.  Due to their long coherence times and optical addressability, solid-state spin defects, such as nitrogen- or silicon-vacancy (NV, SiV) centers, have garnered considerable attention over the past decades as promising qubit platforms \cite{gaebel2006room, PhysRevLett.107.150503, dolde2013room}. However, the generation of robust long-distance entanglement \cite{PhysRevX.3.041023,PhysRevB.99.140403,PRXQuantum.2.040314,andrich2017long} among these qubits has proven experimentally challenging due to environment-driven decoherence, suggesting that reservoir engineering approaches may offer a viable alternative.

A first step in this direction was made by Li \textit{et al.}~\cite{li2023solid}, who developed a formalism describing the quantum many-body dynamics of an ensemble of solid-state spin defects interacting with a common solid-state reservoir. In this work, the authors investigate the strength of dissipation-driven quantum cooperative phenomena, i.e., superradiant and subradiant collective emission \cite{PRXQuantum.3.010201}, by focusing on a stationary $U(1)$-symmetric magnetic bath. Here, we generalize this formalism to non-stationary solid-state baths to then prove that, for magnetic reservoirs, the generation of spin squeezing relies on a time-dependent $U(1)$-symmetry-breaking mechanism.  Our general recipe is summarized in Fig.~\ref{sch}. An ensemble of solid-state qubits with resonance frequency $\omega_q$ interacts with a magnetic bath with spin-wave gap $\Delta_F$ in the regime $\omega_q > \Delta_F$, which allows for the qubit and the bath to exchange energy, in the form of magnetic excitations, at the qubit resonance frequency.
The magnetic bath is, in turn, coupled to an auxiliary mode via a nonlinear interaction that breaks its $U(1)$-symmetry. The auxiliary mode is driven classically at a frequency $\omega_s \approx 2\omega_q$. The nonlinear interaction generates squeezing within the magnetic bath, enabling the simultaneous emission and absorption of correlated magnon pairs between the ensemble and the magnetic bath. This process serves as the source for the generation of steady-state many-body entanglement within the qubit ensemble.

The wide array of magnonic and hybrid magnonic systems exhibiting nonlinearities presents numerous opportunities for implementing our scheme \cite{flebus2021magnonics,zheng2023tutorial}. In this work, to provide concrete estimates for near-future experiments,  we consider the setup illustrated in Fig.~\ref{sch}:  an ensemble of solid-state spin defects is positioned above a ferromagnetic thin film, which is placed on a piezoelectric substrate that generates surface acoustic waves (SAWs) when a voltage is applied. As shown by several experiments \cite{PhysRevB.86.134415, PhysRevLett.106.117601,xu2020nonreciprocal, PhysRevLett.108.176601,maekawa1976surface}, a voltage-driven SAW can induce a periodical strain that couples to the magnetization dynamics of the bath via magneto-elastic interactions. Here, we find that the coupling of the magnetic bath to Love SAWs can induce effective magnon-magnon interactions that are formally analogous to the non-equilibrium two-mode squeezing observed in nonlinear optical systems \cite{boyd2008nonlinear,scully1997quantum}. These non-trivial bath correlations are then transferred to the qubit ensemble, resulting in squeezing dynamics that, for experimentally achievable system parameters,  can be detected through collective spin quadrature measurements. Additionally, we show that further  insights into the entanglement within the system could be gained by probing the relaxation dynamics of the ensemble. Our predictions pave the way for generating stable many-body entangled states within ensembles of solid-state spin defects,  which, in turn, could enable the use of their unique properties in multi-qubit sensing schemes and logic operations.

This paper is organized as follows. In Sec.~\ref{sec1}, we introduce a general framework that describes the dynamics of a qubit array interacting with magnetic field fluctuations emitted by a non-stationary solid-state environment and identify the key ingredients for the generation of steady-state spin squeezing. In Sec.~\ref{sec2}, we outline a recipe for generating non-equilibrium spin squeezing in a magnetic bath to then focus on the concrete example of a SAW-driven Yttrium iron garnet (YIG) thin film. In Sec.~\ref{sec3}, we show that the interaction of an ensemble of NV centers interacting with the bath introduced in Sec.~\ref{sec2} can generate steady-state spin squeezing with the ensemble.  Finally, Sec.~\ref{sec5} we present our conclusion and an outlook. 
\section{General framework}\label{sec1}
We set the stage for our work by deriving a theoretical framework that describes the dynamics of an array of solid-state defects, modelled as two-level systems, interacting with the magnetic field $\mathbf{B}$ generated by a shared, non-stationary solid-state bath. The Hamiltonian of the array of $N$ qubits can be written as  
\begin{equation} \label{eq1}
H= -\frac{1}{2} \sum_{\alpha} \omega_{q} \sigma^z_{\alpha}\\ 
-{\tilde{\gamma}} \sum_{\alpha} (B_{\alpha}^+ \sigma_{\alpha}^- + B_{\alpha}^- \sigma_{\alpha}^+ + B_{\alpha}^z \sigma_{\alpha}^z),
\end{equation}
where $\boldsymbol{\sigma_\alpha}$, with  $\sigma_{\alpha}^{\pm} = (\sigma_{\alpha}^x \pm i \sigma_{\alpha}^y)/2$, labels the quantum spin with resonance frequency  $\omega_{q}$ and gyromagnetic ratio  ${\tilde{\gamma}}$ residing the site $\bm{r}_{\alpha}$, where $\alpha \in \{1, \dots, N\}$.  Here, $B_{\alpha}^{\pm} = B_{\alpha}^x \pm i B_{\alpha}^y$ and $B_{\alpha}^z$ are the components of the stray field interacting with the $\alpha$th quantum spin. The interactions between a solid-state spin defect and the field fluctuations of a nearby solid-state reservoir are much weaker than the characteristic energy scales of the individual systems, and the bath relaxation dynamics considered here are much faster than those of the qubit ensemble. Thus, we can derive the dynamics of the density matrix $\rho_s$ of the qubit array in the Born-Markov approximations \cite{breuer2002theory} as
\begin{align}\label{eq2}
    \frac{d\rho_s(t)}{dt}=-i[H_{eff},\rho_s]+\mathcal{L}(\rho), 
\end{align}
with
\begin{align}
&H_{eff}=\sum_{\alpha,\beta}\sum_{\mu,\bar{\mu}}\sum_{\nu,\bar{\nu}}J^{\mu\nu}_{\alpha\beta}\;\sigma_\alpha^{\bar{\mu}}\sigma_\beta^{\bar{\nu}}\label{ham}, \\
&\mathcal{L}(\rho)=\sum_{\alpha,\beta}\sum_{\mu,\bar{\mu}}\sum_{\nu,\bar{\nu}}\Gamma^{\mu\nu}_{\alpha\beta}\;\left(\sigma_\alpha^{\bar{\mu}}\rho_s\sigma_\beta^{\bar{\nu}}-\frac{1}{2}\{\sigma_\alpha^{\bar{\mu}}\sigma_\beta^{\bar{\nu}},\rho_s\}\right), \label{eq3}
\end{align}
where the sums over different indices are such that $(\mu,\bar{\mu}), (\nu,\bar{\nu})\in \big\{(+,-),(-,+),(z,z)\big\}$. While Eqs.~\eqref{eq2}-\eqref{eq3} are formally identical to the dynamics investigated in Ref.~\cite{li2023solid}, here we generalize them to a non-stationary bath.  The coefficients $J^{\mu\nu}_{\alpha\beta}$ and $\Gamma^{\mu\nu}_{\alpha\beta}$ are given in terms of the two-point magnetic field correlators as 
\begin{align}\label{eq4}
    \frac{J^{\mu\nu}_{\alpha\beta}}{(i\tilde{\gamma}^2/2)}&=
    \int_0^\infty d\tau \; \langle B_\alpha^\mu(t)B_\beta^\nu(t-\tau)\rangle \;e^{i(\omega^\mu_\alpha t+\omega^\nu_\beta (t-\tau))}\nonumber \\ &-\int_0^\infty d\tau \; \langle B_\alpha^\mu(t-\tau)B_\beta^\nu(t)\rangle \;e^{i(\omega^\nu_\beta t+\omega^\mu_\alpha (t-\tau))}\,, 
    \\
   \frac{ \Gamma^{\mu\nu}_{\alpha\beta}}{\tilde{\gamma}^2}&=\int_0^\infty d\tau \; \langle B_\alpha^\mu(t)B_\beta^\nu(t-\tau)\rangle \;e^{i(\omega^\mu_\alpha t+\omega^\nu_\beta (t-\tau))}\nonumber\\&+\int_0^\infty d\tau \; \langle B_\alpha^\mu(t-\tau)B_\beta^\nu(t)\rangle \;e^{i(\omega^\nu_\beta t+\omega^\mu_\alpha (t-\tau))},\label{eq4.1}
\end{align}
where $\omega_{\alpha,\beta}^\pm=\pm\omega_q$, $\omega_{\alpha,\beta}^z=0$. For a stationary bath, i.e., for $t\rightarrow 0$, Eqs.~(\ref{eq4}) and \eqref{eq4.1} reduce, respectively, to the coherent and dissipative inter-qubit couplings derived in Ref.~\cite{li2023solid}. 

In this work, we specialize to a $n$-dimensional magnetically ordered reservoir whose local spin density $\bold{s}(\bm{r},t)$ generates a magnetic stray field 
\begin{align}\label{eq5}
    \bold{B}_\alpha(t)=\int d^n \bm{r} \; \mathcal{D}(\bm{r}_\alpha,\bm{r})\bold{s}(\bm{r},t)\,,
\end{align}
where $\mathcal{D}(\bm{r}_\alpha,\bm{r})$ is the tensorial magnetostatic Green's function~\cite{guslienko2011magnetostatic}. To simplify the analysis, here we assume that the equilibrium orientation of the spin density in the magnetic bath is parallel to the $\hat{z}$ axis. The transverse spin density fluctuations in the magnetic reservoir can be then mapped to the magnon creation, i.e., $ m_{\bold{k}}^\dagger$, and annihilation, i.e., $m_{\bold{k}}$,  operators, by invoking the Holstein-Primakoff transformation~\cite{holstein1940field}. This allows us to rewrite the components of $ \bold{B}_\alpha$ in terms of the magnon operators as proportional to $\int d^2 \bold{k} [a_{\bold{k}}(\bm{\rho}_\alpha) m_{\bold{k}}+b_{\bold{k}}(\bm{\rho}_\alpha) m_{\bold{k}}^\dagger]$, where $a_{\bold{k}}(\bm{\rho}_\alpha)$ and $b_{\bold{k}}(\bm{\rho}_\alpha)$ are coefficients determined by the magnetostatic Green's function\footnote{Note that we neglect the contributions corresponding to two-magnon noise, e.g., $s^z \propto m_{\bold{k}}^\dagger m_{\bold{k}}$, since they are negligible when the qubit resonance frequency $\omega_q$ can be tuned above the spin-wave gap $\Delta_F$ \cite{li2023solid,PhysRevLett.121.187204}.}(see Appendix \ref{A}). As shown in Fig.~\ref{sch}, the qubit frequency is tuned above the spin-wave gap, i.e., $\omega_q > \Delta_F$, allowing for the exchange of real magnons between the qubits and the magnetic bath. On the other hand, when $\omega_q < \Delta_F$,  the Lindbladian (\ref{eq3}) vanishes and the inter-qubit interaction is mediated solely  by virtual magnons. 


In contrast to Ref.~\cite{li2023solid}, here we consider a magnetic reservoir with broken $U(1)$-symmetry, which prevents us from neglecting number non-conserving 
correlations such as $\langle m_{\bold{k}}(t) m_{\bold{k}} 
(t^\prime)\rangle$ and $\langle m_{\bold{k}}^\dagger(t) 
m_{\bold{k}}^\dagger (t^\prime)\rangle$. These correlations — indicative of squeezing in the reservoir — are essential for generating squeezed many-body dynamics in the qubit array. Namely, when $\langle m_{\bold{k}}(t)m_{\bold{k}^\prime}(t^\prime)
\rangle\propto e^{-i\omega_k(t+t^\prime)}$, inter-qubit squeezing interactions of the type
\begin{align}\label{eq5.1}
\frac{\Gamma^{++}_{\alpha\beta}}{(\tilde{\gamma}^2)}&
    \propto\int_0^\infty d^2\bold{k}\; f_{\alpha\beta}(\bold{k})\; e^{-2i(\omega_{\bold{k}}-\omega_q)t} \; \delta (\omega_\bold{k}-\omega_q),
\end{align}
(and similarly its Hermitian conjugate $\Gamma^{--}_{\alpha\beta}$) become stationary if the bath hosts a magnon mode whose dispersion $\omega_{\bold{k}}$ can meet the condition $\omega_{\bold{k}} = \omega_q$.   In this scenario, the Lindbladian (\ref{eq3}) takes the following form: 
\begin{align}\label{eq7}
    \mathcal{L}(\rho)&=\sum_{\alpha,\beta}\Gamma_{\alpha\beta}^{-+}\;\left(\sigma_\alpha^{-}\rho_s\sigma_\beta^{+}-\frac{1}{2}\{\sigma_\alpha^{+}\sigma_\beta^{-},\rho_s\}\right)\nonumber\\
    &+\sum_{\alpha,\beta}{\Gamma}_{\alpha\beta}^{+-}\;\left(\sigma_\alpha^{+}\rho_s\sigma_\beta^{-}-\frac{1}{2}\{\sigma_\alpha^{-}\sigma_\beta^{+},\rho_s\}\right)\nonumber \\
    &+\sum_{\alpha,\beta}{\Gamma}_{\alpha\beta}^{--}\;\left(\sigma_\alpha^{+}\rho_s\sigma_\beta^{+}-\frac{1}{2}\{\sigma_\alpha^{+}\sigma_\beta^{+},\rho_s\}\right)\nonumber \\
    &+\sum_{\alpha,\beta}{\Gamma}_{\alpha\beta}^{++}\;\left (\sigma_\alpha^{-}\rho_s\sigma_\beta^{-}-\frac{1}{2}\{\sigma_\alpha^{-}\sigma_\beta^{-},\rho_s\}\right).
\end{align}
where $\Gamma_{\alpha\beta}^{+-}$ ($\Gamma_{\alpha\beta}^{-+}$)  parameterizes the correlated emission (absorption) of magnons from (by) the array for $\alpha \neq \beta$, while for $\alpha=\beta$  these parameters add up to the single-qubit relaxation rates routinely measured in quantum sensing experiments \cite{du2017control,wang2022noninvasive}. On the other hand, the inter-qubit couplings ${\Gamma}_{\alpha\beta}^{++}$, ${\Gamma}_{\alpha\beta}^{--}$, which can be obtained from Eq.~(\ref{eq4}) under secular approximation \cite{breuer2002theory}, describe the simultaneous generation (and absorption) of a pair of magnons with energy $\omega_{\bold{k}}=\omega_q$  into (from) the reservoir. This mechanism is known to lead to the asymmetric noise reduction and amplification in the collective spin quadratures, and, thus, the generation and stabilization of spin squeezed states in the array \cite{PhysRevLett.110.120402, PhysRevX.12.011015, agarwal1989nonequilibrium, PhysRevA.41.3782,PhysRevA.49.4968}.

It is important to note that, in contrast, the Hamiltonian interaction terms of the form $J_{\alpha\beta}^{++}$ (and, similarly, $J_{\alpha\beta}^{--}$) have the following structure: 
\begin{align}\label{eq12.1}
   \frac{J^{++}_{\alpha\beta}}{(i\tilde{\gamma}^2/2)}&=
    \int_0^\infty d\tau \; \langle B_\alpha^+(t)B_\beta^+(t-\tau)\rangle \;e^{i\omega_q(2t-\tau)}\nonumber \\ &-\int_0^\infty d\tau \; \langle B_\alpha^+(t-\tau)B_\beta^+(t)\rangle \;e^{i\omega_q(2t-\tau)},
\end{align}
where the time-dependent part of the correlations $\langle B_\alpha^+(t)B_\beta^+(t^\prime)\rangle$ is given by $e^{-i\omega_{\bold{k}}(t+t^\prime)}$. It is straightforward to see that the terms inside integral on the  right-hand side (RHS) of Eq.~(\ref{eq12.1}) are identical, which implies  $J_{\alpha\beta}^{++}=J_{\alpha\beta}^{--}=0$.

Furthermore, as discussed in detail in Ref.~\cite{li2023solid}, the typical strength of XY exchange interactions ($J_{\alpha\beta}^{+-}$, $J_{\alpha\beta}^{-+}$)  is considerably higher than that of Ising interactions  ($J_{\alpha\beta}^{zz}$), which allows us to rewrite Hamiltonian (\ref{ham}) as 
\begin{align}\label{eq12}
    H_{eff}=\sum_{\alpha\neq \beta}J_{\alpha\beta}\;\sigma_\alpha^+\sigma_\beta^- \,,
\end{align}
with $J_{\alpha\beta}=J_{\alpha\beta}^{-+}+J_{\alpha\beta}^{+-}$.
In the following, we will show how a nonlinear $U(1)$-symmetry breaking coupling of a magnetic bath to an auxiliary driven mode can be harnessed to generate the array dynamics described by Eqs.~(\ref{eq7}) and (\ref{eq12}).   

\section{Squeezed bath} \label{sec2}

The squeezing dynamics described by Eq.~\eqref{eq7} can be generated by the interaction of a qubit ensemble with a common magnetic reservoir, whose squeezing factors in both quadratures can oscillate resonantly with the qubit resonance frequency $\omega_q$. A key resource for the generation of such non-stationary squeezing correlations in the bath is a nonlinear three-mode interaction of the form $m_\bold{k}m_{\bold{k}^\prime}c^\dagger+m_\bold{k}^\dagger m_{\bold{k}^\prime}^\dagger c$. When the mode $c$ is strongly driven at frequency $\omega_s$ with an undepleted classical field amplitude $g$, that is, $c(t)=ge^{-i\omega_st}$, the interaction reduces to $m_\bold{k}m_{\bold{k}^\prime}g^*+m_\bold{k}^\dagger m_{\bold{k}^\prime}^\dagger g$, \textit{viz.}, a two-mode squeezing interaction of modes $m_\bold{k}$ and $m_{\bold{k}^\prime}$ with an effective interaction strength $g$. This leads to the spontaneous generation of entangled pairs of magnons. Essentially, a quasi-particle with frequency $\omega_s$  is converted into two bath modes $m_\bold{k}$ and $m_{\bold{k}^\prime}$, with frequency $\omega_{\bold{k}}$ (where we assumed $\omega_{\bold{k}}=\omega_{\bold{k}^\prime}$), such that $\omega_s = 2\omega_{\bold{k}}$, and with finite correlations of the type $\langle m_{\bold{k}}(t)m_{\bold{k}^\prime}(t^\prime)
\rangle\propto e^{-i\omega_{\bold{k}}(t+t^\prime)}$. As noted in Sec. \ref{sec1}, when there exists a wavevector $\bold{k_q}$ such that $\omega_{\bold{k_q}}=\omega_q$, these correlations can stabilize the otherwise fast oscillating inter-qubit coupling ${\Gamma}_{\alpha\beta}^{\pm \pm}$ (\ref{eq7}). The latter, in turn, enable the simultaneous exchange of entangled magnon pairs between the bath and the qubit array.

In the following, we demonstrate how the well-established experimental setup depicted in Fig.~\ref{sch} can generate such interactions. Specifically, we consider a ferromagnetic thin film of thickness $L$ mounted on top of a piezoelectric material whose SAW modes can be excited by a remote interdigital transducer (IDT) at frequency $\omega_{s}$ \cite{PhysRevB.86.134415}. The Hamiltonian of the non-interacting ferromagnetic system is given by 
\begin{align}\label{eq8}
    H_F=&\int d^2 \bm{\rho}  \; \Big[-J\bold{s}(\bm{\rho})\cdot\nabla^2\bold{s}(\bm{\rho})-A(\bold{s}(\bm{\rho})\cdot\hat{z})^2-{\gamma}{B}_0\bold{s}(\bm{\rho})\cdot\hat{z}\Big],
\end{align} where $J$ and $A$ parameterize, respectively, the strength of the symmetric (Heisenberg-like) exchange interaction and the uniaxial anisotropy, $\bold{B}_0$ is the applied static magnetic field, while $\bold{s}(\bm{\rho})$ here denotes the surface spin density. Here onward, we denote three-dimensional Cartesian coordinates by \(\boldsymbol{r}\) and represent the two-dimensional coordinates within the plane of the magnetic film using \(\boldsymbol{\rho}\), i.e., \(\boldsymbol{r} = (\boldsymbol{\rho}, z)\). The magnetic excitations are coupled to the elastic displacement via magnetoelastic interactions, which, for a cubic solid with lattice constant $a_0$, can be written as \cite{gurevich2020magnetization, PhysRevB.95.144420} 
\begin{align}\label{eq9}
H_{int}=\frac{nL}{s^2}\int d^2\bm{\rho}\sum_{\upsilon\eta}[\mathcal{B}_{\upsilon\eta}{s}^\upsilon(\bm{\rho}){s}^\eta(\bm{\rho})]\mathcal{E}_{\upsilon\eta}.
\end{align}
Here, $s$ is the saturation surface spin density, $n=1/a_0^3$, $\mathcal{B}_{\upsilon\eta}$ are the magneto-elastic coupling strengths and $\mathcal{E}_{\upsilon\eta}$ is the elastic strain defined as $\mathcal{E}_{\upsilon\eta}=\frac{1}{2}\Big(\frac{\partial u_\upsilon}{\partial \eta}+\frac{\partial u_\eta}{\partial \upsilon}\Big)$, $\upsilon, \eta \in \{x,y,z\}$ where $\bold{u}$ describes the displacement field of the SAW. In order to generate the target quadrature squeezing, we consider a Love SAW with frequency $\omega_s$ and wavevector $\bold{k_s}$ propagating along the $\hat{x}$ direction with polarization along $\hat{y}$, as shown in Fig.~\ref{sch}. Since the typical experimental microwave drive powers can generate a displacement amplitude in the nanometer range \cite{PhysRevB.86.134415}, which is significantly larger than the femtometer domain of the SAW zero-point motion amplitude \cite{PhysRevX.5.031031}, one can describe the SAW  as a classical displacement field with amplitude $u_0$, i.e., $u_y=u_0\sin(k_sx-\omega_s t)$. The only non-zero strain components of the Love SAW are $\mathcal{E}_{xy}$ and $\mathcal{E}_{yx}$.



Invoking the Holstein-Primakoff (HP) transformation \cite{holstein1940field}, from Eqs.~(\ref{eq8}) and~(\ref{eq9}) we can recast the bath Hamiltonian $H_{tot}=H_{F}+H_{int}$ as (see Appendix \ref{B} for details)
\begin{align}\label{eq10}
    H_{tot}=\int d^{2}\bold{k}\left[\omega_{{k}}m_{\bold{k}}^\dagger m_{\bold{k}}+2g\cos\left(\omega_s t\right)\left(m_{\bold{k}}m_{-\bold{k}+\bold{k}_s}-m_{\bold{k}}^\dagger m_{-\bold{k}+\bold{k}_s}^\dagger\right)\right],
\end{align}
 where $\omega_{{k}}=Dk^2+2As+\gamma B_0$ is the spin-wave dispersion, $\gamma$ the gyromagnetic ratio, $D=Js$ the spin stiffness, $g\approx g_{\bold{k_q}}=-i{n L}{\mathcal{B}_{xy}}\mathcal{E}_{xy}/2s$ is the static amplitude of the classically-driven field and the bosonic operators $m_{\bold{k}}$ obey the commutation relation $[m_{\bold{k}},m_{\bold{k}^\prime}^\dagger]=(2\pi)^2\delta^2(\bold{k}-\bold{k}^\prime)$, where $\delta^2(\bold{k}-\bold{k}^\prime)$ is the Dirac delta function. Generally, magnons interact with their phononic environment even in the absence of external drives \cite{PhysRevB.95.144420}. However, in the setup we consider, the stationary magneto-elastic interactions are much weaker than the coupling to the classically-driven Love wave and do not play a substantial role in our analysis.
 
The second term on the RHS of Eq.~(\ref{eq10}) describes the conversion of a phonon with frequency $\omega_s$  into two magnons with frequency $\omega_{{k}}$ (and \textit{vice versa}), i.e., a process that breaks the $U(1)$-symmetry of the isolated magnon reservoir. As a result, this interaction mediates a two-mode non-equilibrium squeezing that allows for the generation of a squeezed state formally analogous to the squeezed vacuum generated in an optical system through a nonlinear medium \cite{boyd2008nonlinear, scully1997quantum}. 
This non-equilibrium bath squeezing might be probed experimentally by measuring the quadrature of the collective spin in a direction perpendicular to its equilibrium direction, similar to the techniques used in atomic ensembles \cite{hosten2016quantum,leroux2010implementation}.

For a narrow range of $\bold{k}$ centered around the value for which $\omega_{{k}}= \omega_{s}/2$, where $\bold{k}$ lies within the band of modes exhibiting significant squeezing relative to the central $\bold{k}$ mode, the zero-temperature magnon bath correlations can be written as  
\begin{align} 
    &\langle m_{\bold{k}}(t)m_{\bold{k}^\prime}(t^\prime)\rangle=(2\pi)^2M_k e^{-i\omega_k(t+t^\prime)}\;\delta^2({\bold{k}+\bold{k}^\prime}),\label{eq11}
    \\
    &\langle m_{\bold{k}}^\dagger(t)m_{\bold{k}^\prime}^\dagger(t^\prime)\rangle=(2\pi)^2M_k^* e^{i\omega_k(t+t^\prime)}\;\delta^2({\bold{k}+\bold{k}^\prime}),\label{eq11.1}
    \\
    &\langle m_{\bold{k}}^\dagger(t)m_{\bold{k}^\prime}(t^\prime)\rangle=(2\pi)^2N_ke^{i\omega_k(t-t^\prime)}\;\delta^2({\bold{k}-\bold{k}^\prime}),\label{eq11.2}
    \\
    &\langle m_{\bold{k}}(t)m_{\bold{k}^\prime}^\dagger(t^\prime)\rangle=(2\pi)^2(N_k+1)e^{-i\omega_k(t-t^\prime)}\;\delta^2({\bold{k}-\bold{k}^\prime}),\label{eq11.3}
\end{align} 
where the contributions from $\bold{k_s}$ in the Dirac delta functions are neglected (see Appendix \ref{B} for details). Here, $N_k=\sinh^2(r_k)$, $M_k=-\cosh(r_k)\sinh(r_k)e^{i\phi}$, $\phi=-i\text{log}({g}/{|g|})$, and the squeezing parameter $r_k$, for $\omega_k\approx\omega_s/2$, is given by  $r_{k_q}=\frac{1}{2}\text{arctanh}\left({|g|}/{\bar{\Delta}}\right)$, where $\bar{\Delta}$ denotes the bandwidth of the squeezed vacuum. We work in the stable domain of system parameters, i.e., $|g|<\bar{\Delta}$ for which $r_{k_q}\approx{|g|}/{2\bar{\Delta}}$. 
Here the magnon correlations (\ref{eq11}) and ~(\ref{eq11.1}) can stabilize the squeezing inter-qubit squeezing interactions, as shown by  Eq.~(\ref{eq5.1}). Effectively, a phonon with frequency $\omega_s$ generates a correlated pair of magnons at frequency $\omega_k$, resulting in squeezing correlations within the magnon bath. When the qubit system has a frequency $\omega_q>\Delta_F$ such that $\omega_q \approx \omega_s/2$, these squeezing correlations are transmitted to the qubit array, as we will show in detail in the next Section.

\section{Steady-state spin squeezing}\label{sec3}
We now focus on tailoring the master equation (\ref{eq2}) with Lindbladian (\ref{eq7}) and effective Hamiltonian (\ref{eq12}) to describe an array of NV centers with lattice constant $a$ interacting dissipatively with the non-equilibrium squeezed ferromagnetic reservoir introduced in Sec.~\ref{sec2}. An NV center is a spin-triplet system whose degeneracy of the $m_s=\pm1$ is lifted by the applied bias magnetic field $\bold{B}_0$. In other words, each qubit can then be modelled as a two-level system with a frequency $ \omega_{q} = \Delta_0 - \tilde{\gamma} B_0 $, where $ \Delta_0 = 2.87 \, \text{GHz}$  represents the zero-field splitting of the NV center. This approximation is valid as long as the thermal energy of the system is much less than the energy of the higher NV center transition, i.e., $ k_B T \ll \Delta_0 + \tilde{\gamma} B_0$, which is compatible with the quantum regime $T\rightarrow 0$ we are working in. 
To guarantee the emergence of squeezing interactions $\propto \Gamma_{\alpha\beta}^{\pm\pm}$, we take the NV resonance frequency $\omega_q$ to be within the magnon continuum, i.e., $\omega_q>\Delta_F$ and the SAW drive frequency is chosen to satisfy the frequency matching condition  $\omega_s= 2\omega_q$.

\begin{figure*}
\captionsetup{justification=raggedright,singlelinecheck=false}
 \centering
   \includegraphics[scale=0.55]{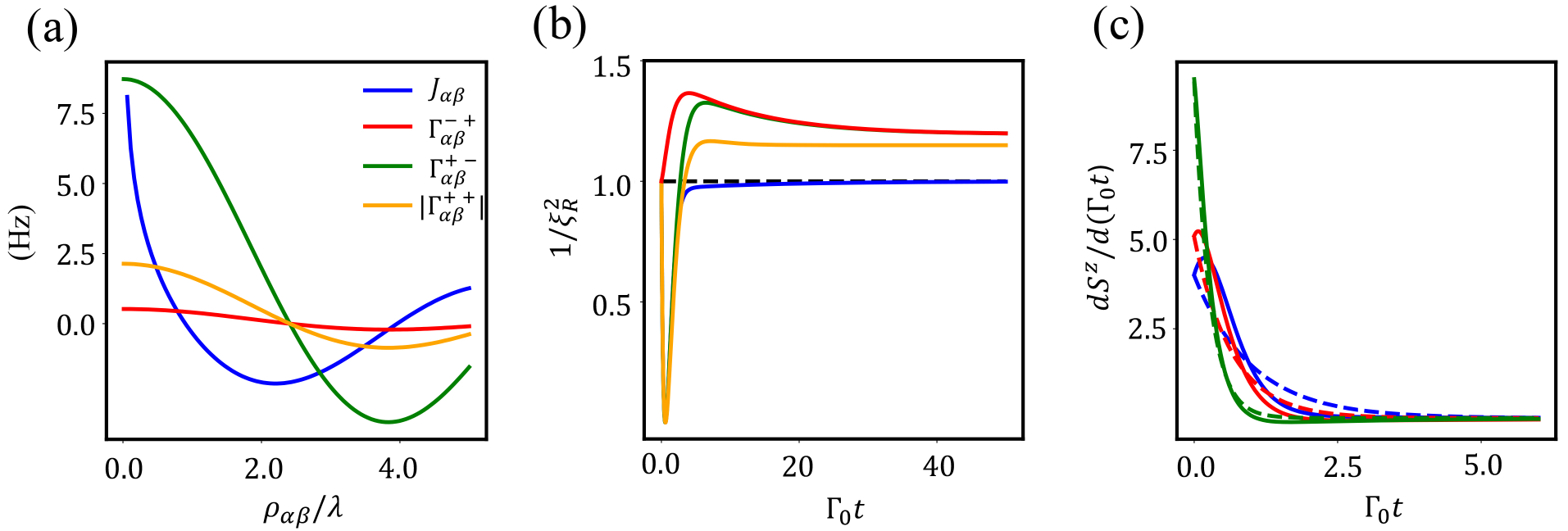}
\caption{(a) Dependence of the coherent and dissipative interqubit couplings defined in Eqs.~(\ref{eq13})-(\ref{eq16}) on the inter-qubit separation $\rho_{\alpha\beta}$ (in units of the wavelength $\lambda$). The parameters used are for an ensemble of NV-center spins interacting via a YIG thin film driven into the non-equilibrium squeezed state detailed in the main text. (b) Time evolution of the inverse of Wineland's squeezing parameter $1/\xi_R^2$~\eqref{wine}  for  $N=2$ NV centers. The blue and green curves correspond to the squeezing dynamics of an ensemble with lattice constant $a/\lambda=0.5$ initialized in the fully excited state with, respectively, squeezing parameter $r_{k_q}=0$ and $r_{k_q}=0.25$, whereas the orange curve corresponds to an ensemble with lattice constant $a/\lambda=1$ and squeezing parameter $r_{k_q}=0.25$. The red curve represents the squeezing dynamics of the NV ensemble initialized in the ground state with squeezing parameter $r_{k_q}=0.25$ and lattice constant $a/\lambda=0.5$. (c) The relaxation dynamics of 
$N=4$ NV centers are compared for a non-interacting ensemble (dotted lines with $a/\lambda>>1$) and a correlated array (solid lines with lattice constant $a/\lambda=0.4$). The blue, red and green curves correspond to the squeezing parameter $r_{k_q}=0, 0.5, 1$, respectively. (b,c) The time $t$ is expressed in terms of the dimensionless quantity $\Gamma_{0}t$, where $\Gamma_{0}$ is the relaxation rate of a noninteracting array.}
\label{squee}
\end{figure*}
When the distance $d$ between the NVs and the magnetic film is much smaller than the characteristic wavelength $\lambda=\sqrt{{D}/{\left(\omega_q-\Delta_F\right)}}$, the inter-qubit couplings can be written as 
\begin{align}\label{eq13}
    J_{\alpha\beta}/\nu&=-\frac{\pi}{2}\frac{\omega_q-\Delta_F}{\Delta_0}Y_0(\rho_{\alpha\beta}/\lambda),
    \\
    \Gamma_{\alpha\beta}^{-+}/\nu&={\pi}\frac{\omega_q-\Delta_F}{\Delta_0}N_{k_q}J_0(\rho_{\alpha\beta}/\lambda), \label{eq14}
    \\
    \Gamma_{\alpha\beta}^{+-}/\nu
    &={\pi}\frac{\omega_q-\Delta_F}{\Delta_0}(N_{k_q}+1)J_0(\rho_{\alpha\beta}/\lambda),\label{eq15}\\
    \Gamma_{\alpha\beta}^{++}/\nu&={\pi}\frac{\omega_q-\Delta_F}{\Delta_0}M_{k_q}^*J_0(\rho_{\alpha\beta}/\lambda),\label{eq16}
    \\
    \Gamma_{\alpha\beta}^{--}/\nu&={\pi}\frac{\omega_q-\Delta_F}{\Delta_0}M_{k_q}J_0(\rho_{\alpha\beta}/\lambda).\label{eq17}
\end{align}
 Here,  $\rho_{\alpha\beta}=|\bm{\rho}_\alpha-\bm{\rho}_\beta|$ is the distance between the $\alpha${th} and the $\beta$\text{th} quantum spins, and $\nu={\pi h^3(\gamma\tilde{\gamma})^2s\Delta_0}/{D^2}$ is a characteristic frequency controlling the strength of the inter-qubit interactions. For concreteness, here we consider a YIG thin film with a thickness of $L = 20$ nm, $a_0=12.3$ $\AA$, $D =5.1\times10^{-28}$ erg$\cdot$cm$^{2}$, $s=1.2\times10^{-10}$ G$^2\cdot$cm$\cdot$s, zero field gap $2As=3.6\times 10^{-18}$ $\text{erg}\cdot \text{cm}^2$ and $\mathcal{B}_{xy}=1988$ GHz \cite{gurevich2020magnetization, PhysRevB.95.144420}. In order to realize an ensemble of qubits with long-range dissipative interactions, one needs to maximize $\lambda$, which can be achieved by minimizing $\omega_q - \Delta_F$. For $B_0 \approx 40$ mT, one finds $\omega_q - \Delta_F \approx 100$ MHz, which implies $\lambda \approx 277$ nm and $\nu\approx75$ Hz. The squeezing parameter $r_{k_q}$ depends on the nonlinear interaction strength $g$, which, in turn, is a function of $\mathcal{E}_{xy}$ controlled by the power ($P_{SAW}$) of the SAW drive, i.e., $\mathcal{E}_{xy}\propto\sqrt{P_{SAW}}$. For SAW drive powers close to 80 mW and velocity of the SAW $\approx6000$ m/s, one finds $\mathcal{E}_{xy}\approx10^{-4}$ \cite{PhysRevB.86.134415}. The corresponding nonlinear coupling strength of $g\approx0.1$ MHz and $\bar{\Delta}\approx 0.25$ MHz can generate a squeezing $r_{k_q} \approx 0.2$, which yields $N_{k_q} \approx 0.04$ and $|M_{k_q}| \approx 0.2$, where $N_{k_q}$ represents the average magnon number at temperature $T=0$. Figure~\ref{squee} (a) shows the dependence of the coherent and dissipative inter-qubit interactions described by Eqs.~(\ref{eq13})-(\ref{eq17}) on the inter-qubit separation $\rho_{\alpha\beta}$. As one would expect, when $r_k \rightarrow 0$, Eqs.~(\ref{eq14}), (\ref{eq16}), and (\ref{eq17}) vanish, leaving only the interaction $\Gamma_{\alpha\beta}^{+-}$ describing correlated emission, which is responsible for the emergence of superradiant and subradiant collective dynamics~\cite{li2023solid}.  
Given that $N_{k_q} \ll 1$, the magnitude of $\Gamma_{\alpha\beta}^{-+}$ is much smaller than that of the other interaction terms. That is, the rate at which the NV array absorbs single magnons from the bath is considerably weaker than the rates at which it absorbs magnon pairs and emits both single magnons and magnon pairs into the bath. Nonetheless, the vacuum fluctuations of the bath can play a role analogous to that of a finite thermal magnon population in the emergence of superradiant and subradiant cooperative dynamics, as we discuss in detail later on. Finally, we note that the structure of the Lindbladian (\ref{eq7}) with the inter-qubit interactions given by Eqs.~(\ref{eq14})-(\ref{eq17}) is in one-to-one correspondence with squeezed vacuum reservoirs in quantum-optical systems \cite{breuer2002theory,PhysRevA.41.3782}.  In analogy with the latter~\cite{breuer2002theory}, we can introduce a jump operator $C_\alpha=\cosh(r_{k_q})\sigma_\alpha^-+\sinh(r_{k_q})e^{i\phi}\sigma_\alpha^+$ that allows to recast Eq.~(\ref{eq7}) as 
\begin{align}\label{eq18}
    \mathcal{L}(\rho)={\pi}\frac{\omega_q-\Delta_F}{\Delta_0}\sum_{\alpha,\beta}J_0(\rho_{\alpha\beta}/\lambda)\left[\frac{1}{2}\{C_\alpha^{\dagger}C_\beta,\rho_s\}-C_\alpha\rho_sC_\beta^{\dagger}\right],
\end{align}
which shows that, in the long-time limit, the system decays into the Gibbs ensemble in the long-time limit, exhibiting stable spin squeezing. 

We now proceed to explore the generation of steady-state spin squeezing in an array of $N$ qubits whose dynamics is described by Eqs.~(\ref{eq7}) and~(\ref{eq12}), with coherent and dissipative exchange interactions given by Eqs.~(\ref{eq13})-(\ref{eq17}).  To quantify the degree of spin squeezing, we use the Wineland squeezing parameter 
\begin{align}\label{wine}
\xi_R^2=N\frac{\langle\Delta \bold{S}_{\perp}^2\rangle}{\langle \bold{S}\rangle^2},
\end{align}
where $\langle\Delta \bold{S}_{\perp}^2\rangle$ denotes the minimum variance in the plane perpendicular to the mean spin direction and the collective spin operators are defined as $S^{\eta}=\sum_{i=1}^N\sigma_i^\eta$. A value of $\xi_R^2<1$ indicates that the collective spin states are squeezed, which, in the case of an ensemble of spin-$1/2$ particles, signals finite multipartite entanglement \cite{sorensen2001many}.

We start our analysis by numerically simulating the dynamics of the system for  $N=2$ qubits. We use a scaled time $\Gamma_0t$ to simulate the system's time evolution, where $\Gamma_0 \equiv \Gamma_{\alpha\alpha} \approx 8.2$ Hz is the relaxation rate of an isolated NV center interacting  with the magnetic bath. The blue curve in Fig.~\ref{squee}(b) shows the temporal evolution of  (the inverse of) the spin squeezing parameter (\ref{wine}) for an ensemble initialized in the excited state with $r_{k_q}=0$ and $a/\lambda=0.5$. As one would expect, the system does not exhibit any squeezing when the bath is not driven into a squeezed state, i.e.,   $1/\xi_R^2\rightarrow1$ in the long-time limit (blue curve).  In contrast, for $r_{k_q}=0.25$ and $a/\lambda=0.5$, we find that the two qubits reach an entangled steady state, i.e., $1/\xi_R^2>1$ in the long-time limit (green curve). As $a/\lambda$ increases, the strength of the inter-qubit interaction diminishes, resulting in a weaker steady-state squeezing (orange curve).

The red curve in Fig.~\ref{squee}(b) shows the time of evolution of $1/\xi_R^2$  for an ensemble initialized in the ground state, with  $r_{k_q}=0.25$ and $a/\lambda=0.5$. Unlike the all-excited state,  the all-down state maintains $1/\xi_R^2>1$  at all times. The discrepancy in the short-time evolution of the red and green curves arises from their distinct initializations: while the ensemble initialized in the ground state rapidly evolves into a squeezed state through its interaction with the bath, the excited ensemble decays into the squeezed environment before exhibiting spin squeezing. Nonetheless, ultimately both ensembles converge to a steady state with the same degree of spin squeezing, confirming that the long-time behavior of the ensemble is independent of the initial state.

Figure~\ref{squee}(c) shows the time evolution of the collective relaxation rate for an array of $N = 4$ qubits initialized in the excited state, with varying $r_{k_q}$. In the absence of squeezing ($r_{k_q} = 0$), the collective relaxation rate of a correlated ensemble (solid blue line) exhibits a characteristic enhancement followed by a subradiant tail, in contrast to the exponentially decaying dynamics of an uncorrelated ensemble of the same size (dashed blue line). However, as the squeezing parameter $r_{k_q}$ increases, the bath becomes more populated, leading to a suppression of the inter-qubit correlations. When the squeezing of the bath reaches $r_{k_q} = 1$ (green curve), the system undergoes collective relaxation dynamics nearly identical to that of an uncorrelated ensemble, suggesting that the collective relaxation dynamics itself  could serve as a qualitative indicator of the presence or absence of squeezing in the reservoir. For instance, in the setup we consider, non-equilibrium squeezing in the bath could be detected by probing the relative suppression or amplification of superradiant and subradiant dynamics when the SAW voltage drive is switched, respectively, on or off.

\section{Conclusions}\label{sec5}
In this work, we develop a theoretical framework to describe the many-body quantum dynamics of a spin qubit array coupled to a nonstationary solid-state bath. Specializing to a magnetically ordered reservoir, we identify a key ingredient for generating steady-state spin squeezing within the ensemble: the presence of nonequilibrium squeezed states in the bath, i.e., correlated magnon pairs at the qubit resonance frequency. We show that such states can be engineered through a nonlinear, $U(1)$-symmetry-breaking interaction between magnons and a driven auxiliary mode. 

Focusing for concreteness on the interaction between a YIG  thin film driven by surface acoustic waves and an ensemble of NV centers, we demonstrate that, within experimentally feasible parameters, the bath-qubit coupling can drive the system into a steady-state spin-squeezed phase. We also find that the emergence of spin squeezing—tunable via an external voltage—can be detected through collective spin measurements, circumventing the need for individual qubit addressability. While we focus on a specific experimental platform, our results highlight the broader potential of magnetic reservoirs for engineering nonequilibrium quantum states. The demonstrated control over magnon squeezing via classical drives extends beyond SAWs, with alternative implementations, such as  parametric magnon excitation via microwave photonic pumping \cite{PhysRevLett.106.216601}. Thus, the rich landscape of nonlinear magnon interactions and external control protocols unlocks a wide array of possibilities \cite{flebus2021magnonics,zheng2023tutorial,flebus20242024} for implementing and extending our framework that future investigations should address.

\section{ACKNOWLEDGEMENTS}
The authors thank A. A. Garcia, A. Clerk, L. Viola, G. S. Agarwal, V. Flynn and X. Li for helpful discussions, and acknowledge support from DOE under Award No. DE-SC0024090.

\appendix
\section{\MakeUppercase{Magnetostatic stray field}} \label{A}
This Appendix details the derivation of the magnetostatic field 
$\bold{B}_\alpha(t)\equiv\bold{B}\left(\bm{r}_\alpha, t\right)$ (\ref{eq5})  in terms of the bath magnon operators.  In a magnetically ordered medium free of charge, fluctuations of the three-dimensional spin density $\mathbf{S}$ generate a magnetic field $ \mathbf{B} = {\nabla} \phi_m$ at position $\bm{r}$, where
\begin{align}
\phi_m\left(\bm{r},t\right) = -{4\pi \gamma}  \int d^3\bm{r}' \nabla' G\left(\bm{r}, \bm{r}'\right) \cdot \mathbf{S}\left(\bm{r}',t\right), \label{a1}
\end{align}
is the magnetic scalar potential and $ G\left(\bm{r}, \bm{r}'\right) $ represents the magnetostatic Green's function. The corresponding magnetic field components can be expressed as
\begin{align}
    B^\upsilon
    (\bm{r},t) = 4\pi\gamma \int d^3\bm{r}' [\partial_\upsilon \partial_\eta' G\left(\bm{r}, \bm{r}'\right)] S^\eta\left(\bm{r}',t\right).\label{a2}
\end{align}
where, for a magnetic film of thickness $L$, the magnetostatic Green function is given by
\begin{align}
    G\left(\bm{r}, \bm{r}'\right) = \frac{1}{2\pi} \int d^2\bold{k} \frac{e^{-k|z-z'|}}{k} e^{i\mathbf{k} \cdot \left(\bm{\rho} - \bm{\rho}'\right)}.\label{a3}
\end{align}
Plugging Eq. (\ref{a3}) into Eq. (\ref{a2}), the magnetic field components in the plane of the qubits, $\left(\text{ i.e.,}\; z - z' = d \right)$ as a function of the thin film coordinates and two-dimensional spin-density $\mathbf{s}=\mathbf{S}L$ can be written as
\begin{align}
    B^\upsilon\left(\bm{\rho},t\right) = 4\pi\gamma \int d^2\bm{\rho}' D_{\upsilon\eta}\left(\bm{\rho} - \bm{\rho}'\right) s^\eta\left(\bm{\rho}',t\right), \label{a4}
\end{align}
where
\begin{align}
    D_{\upsilon\eta}\left(\bm{\rho} - \bm{\rho}'\right) = \partial_\upsilon \partial_\eta' G\left(\bm{\rho} - \bm{\rho}'\right). \label{a5}
\end{align}
Equation (\ref{a5}) can be rewritten in the Fourier domain as
\begin{align}
    D_{\upsilon\eta}\left(\bm{\rho} - \bm{\rho}'\right) = \frac{-1}{2\pi} \int d^2\bold{k} \, \tilde{D}_{\mathbf{k}} \, e^{i\mathbf{k} \cdot \left(\bm{\rho} - \bm{\rho}'\right)} e^{-kd}, \label{a6}
\end{align}
where 
\begin{align}
    \tilde{D}_{\mathbf{k}} = \begin{pmatrix}
    \cos^2\left(\phi_{\mathbf{k}}\right) & \cos\left(\phi_{\mathbf{k}}\right)\sin\left(\phi_{\mathbf{k}}\right) & i\cos\left(\phi_{\mathbf{k}}\right) \\
    \cos\left(\phi_{\mathbf{k}}\right)\sin\left(\phi_{\mathbf{k}}\right) & \sin^2\left(\phi_{\mathbf{k}}\right) & i\sin\left(\phi_{\mathbf{k}}\right) \\
    i\cos\left(\phi_{\mathbf{k}}\right) & i\sin\left(\phi_{\mathbf{k}}\right) & -1
    \end{pmatrix}. \label{a7}
\end{align}
Using Eqs. (\ref{a6}), (\ref{a7}) in Eq. (\ref{a4}), the magnetic field components generated by the thin film at the location of the $\alpha$th qubit are 
\begin{align}
    B^-_\alpha\left(t\right) &= \gamma \int d^2\rho \int \frac{d^2\bold{k}}{\left(2\pi\right)^2} \big[a_{\mathbf{k}} s^-\left(\bm{\rho}, t\right) + b_{\mathbf{k}} s^+\left(\bm{\rho}, t\right) \nonumber \\
    &\quad + c_{\mathbf{k}} s^z\left(\bm{\rho}, t\right)\big] e^{i\mathbf{k} \cdot \left(\bm{\rho}_\alpha - \bm{\rho}\right)} e^{-kd}, \label{a8}\\
    B^+_\alpha\left(t\right) &= \gamma \int d^2\rho \int \frac{d^2\bold{k}}{\left(2\pi\right)^2} \big[a_{\mathbf{k}}^* s^+\left(\bm{\rho}, t\right) + b_{\mathbf{k}}^* s^-\left(\bm{\rho}, t\right) \nonumber \\
    &\quad + c_{\mathbf{k}}^* s^z\left(\bm{\rho}, t\right)\big] e^{-i\mathbf{k} \cdot \left(\bm{\rho}_\alpha - \bm{\rho}\right)}e^{-kd},\label{a9} \\
    B^z_\alpha\left(t\right) &= \gamma \int d^2\rho \int \frac{d^2\bold{k}}{\left(2\pi\right)^2} \big[d_{\mathbf{k}} s^-\left(\bm{\rho}, t\right) + e_{\mathbf{k}} s^+\left(\bm{\rho}, t\right) \nonumber \\
    &\quad + f_{\mathbf{k}} s^z\left(\bm{\rho}, t\right)\big] e^{i\mathbf{k} \cdot \left(\bm{\rho}_\alpha - \bm{\rho}\right)}e^{-kd}, \label{a10}
\end{align}
where $s^{\pm} = (s^x \pm i s^y)/2$ and the coefficients are defined as:
\begin{align}
    a_{\mathbf{k}} &= -2\pi k e^{-kd} [\cos\left(\phi_{\mathbf{k}}\right) - i\sin\left(\phi_{\mathbf{k}}\right)\cos\left(\theta\right) - \sin\left(\theta\right)] e^{i\phi_{\mathbf{k}}}, \label{a11} \\
    b_{\mathbf{k}} &= -2\pi k e^{-kd} [\cos\left(\phi_{\mathbf{k}}\right) - i\sin\left(\phi_{\mathbf{k}}\right)\cos\left(\theta\right) - \sin\left(\theta\right)] e^{-i\phi_{\mathbf{k}}}, \label{a12} \\
    c_{\mathbf{k}} &= -2\pi i k e^{-kd} [\cos\left(\phi_{\mathbf{k}}\right) - i\sin\left(\phi_{\mathbf{k}}\right)\cos\left(\theta\right) - \sin\left(\theta\right)], \label{a13} \\
    d_{\mathbf{k}} &= -2\pi k e^{-kd} [\sin\left(\phi_{\mathbf{k}}\right)\sin\left(\theta\right) + i\cos\left(\theta\right)] e^{i\phi_{\mathbf{k}}}, \label{a14} \\
    e_{\mathbf{k}} &= -2\pi k e^{-kd} [\sin\left(\phi_{\mathbf{k}}\right)\sin\left(\theta\right) + i\cos\left(\theta\right)] e^{-i\phi_{\mathbf{k}}}, \label{a15} \\
    f_{\mathbf{k}} &= -2\pi i k e^{-kd} [\sin\left(\phi_{\mathbf{k}}\right)\sin\left(\theta\right) + i\cos\left(\theta\right)]. \label{a16}
\end{align}
As discussed in the main text, we ignore $s^z$ and $B^z$ terms due to their negligible contributions.  Finally, Eqs. (\ref{a8})-(\ref{a10}) are rewritten in terms of the magnon operators as: 
\begin{align}
    B^-_\alpha\left(t\right) &= \gamma \sqrt{\frac{s}{2}} \int \frac{d^2\bold{k}}{\left(2\pi\right)^2} \big[a_{\mathbf{k}} m_{\mathbf{k}}^\dagger\left(t\right)+ b_{\mathbf{k}} m_{\mathbf{k}}\left(t\right)\big] e^{i\mathbf{k} \cdot \bm{\rho}_\alpha}e^{-kd}, \label{a17} \\
    B^+_\alpha\left(t\right) &= \gamma \sqrt{\frac{s}{2}} \int \frac{d^2\bold{k}}{\left(2\pi\right)^2} \big[a_{\mathbf{k}}^* m_{\mathbf{k}}\left(t\right) + b_{\mathbf{k}}^* m_{\mathbf{k}}^\dagger\left(t\right)\big] e^{-i\mathbf{k} \cdot \bm{\rho}_\alpha}e^{-kd}, \label{a18}
\end{align}
where we used \cite{holstein1940field}
\begin{align}\label{A16}
s^-(\bm{\rho}) = \int \frac{d^2\bold{k}}{(2\pi)^2} s_{\bold{k}}^- e^{i \bold{k} \cdot \bm{\rho}}, \quad \text{and} \quad s_{\bold{k}}^- = \sqrt{\frac{s}{2}} m_{\bold{k}}^\dagger. 
\end{align}  
\section{\MakeUppercase{Squeezed magnons}}\label{B}
In this Appendix, we outline the derivation of the two-mode magnon squeezing interaction~(\ref{eq10}) and magnon correlations ~(\ref{eq11})-(\ref{eq11.3}) arising from the nonlinear three-mode magnon-phonon interaction, with the phonon mode being classically driven by surface acoustic waves. 
 We begin by considering the spin-phonon interaction described by Eq.~(\ref{eq9})
where the phonons are quantized as
\begin{equation}\label{B2}
u_\upsilon(\bm{\rho}, t) = \sum_{\bold{k}, \lambda} \epsilon_\upsilon
^\lambda \sqrt{\frac{\hbar}{2 \rho V \omega_{\lambda \bold{k}}}} \left( b_{\bold{k}\lambda} e^{i \bold{k} \cdot \bm{\rho}} + \text{h.c.} \right),
\end{equation}
where $\epsilon_\upsilon^\lambda$ are components of the unit polarization vector of the phonon mode, and $b_{\bold{k}\lambda}$, $b_{\bold{k}\lambda}^\dagger$ are the annihilation (creation) operators with mass density $\rho$ and the mode volume $V$.  Here, \( \lambda = 1, 2, 3 \) represents transverse and longitudinal polarizations of the acoustic phonons, and $\omega_{\lambda \bold{k}}$  the resonance frequencies. Focusing on a Love SAW driving the magnetic film, we set $\bold{k}_s = k_s \hat{x}$, and $\epsilon^\lambda(\bold{k}_s) = (0, 1, 0)$. The strain components of the SAW are given by
\begin{equation}\label{B3}
\mathcal{E}_{\upsilon \eta}(\bm{\rho}) = \frac{1}{2} \sqrt{\frac{\hbar}{2 \rho V \omega_{k_s}}} 
\left[
\epsilon_\upsilon^\lambda (i {k_{s}}_\eta) b_{\bold{k}_s} e^{i \bold{k}_s \cdot \bm{\rho}} 
- \epsilon_\eta^\lambda (i {k_{s}}_\upsilon) b_{\bold{k}_s} e^{i \bold{k}_s \cdot \bm{\rho}} 
\right].
\end{equation}
Here, the mode $b_{\bold{k}_s}$ is driven classically, which allows allows  to replace the operator $b_{\bold{k}_s}$ with its classical form, i.e., $b_{\bold{k}_s} = i \mathcal{A} f(t)$, where $\mathcal{A}$ and $f(t)$ are taken to be a (real) number representing the classical amplitude of $b_{\bold{k}_s}$ (generated by the external drive) and a (real) function of time, respectively.
In the geometry we consider, the only nonzero components of the strain tensor $\mathcal{E}_{\upsilon \eta}$ are $\mathcal{E}_{xy} = |\mathcal{E}_{xy}| f(t) e^{-i \bold{k}_s \cdot \bm{\rho}}$; $\mathcal{E}_{yx} = |\mathcal{E}_{xy}| f(t) e^{i \bold{k}_s \cdot \bm{\rho}}$, where $|\mathcal{E}_{xy}| = \frac{1}{2} \sqrt{{\hbar}/{2 \rho V \omega_{k_s}}} {k_s} \mathcal{A}$. Thus,  invoking Eq.~(\ref{A16}), we can rewrite Eq.~(\ref{eq9}) as
\begin{equation}\label{B4}
H_{\text{int}} = \frac{nL}{s} \mathcal{B}_{xy} \frac{|\mathcal{E}_{xy}|}{8i} f(t) \sum_{\pm} \int \frac{d^2\bold{k}}{(2\pi)^2} \left[ m_{\bold{k}} m_{-\bold{k}\pm \bold{k}_s} - m_{\bold{k}}^\dagger m_{-\bold{k} \pm \bold{k}_s}^\dagger \right].
\end{equation}
When the applied signal $f(t)$ has carrier frequency $\omega_s$, such that $2\omega_{k_q} \approx \omega_s$, for a small continuum of $\bold{k}$ around $\bold{k}_q$, the Hamiltonian~(\ref{B4}) in the interaction picture can be written as:
\begin{align}\label{B5}
H_{\text{int}} = 
\int \frac{d^2\bold{k}}{(2\pi)^2} \frac{g_\bold{k}}{2}\left[ m_{\bold{k}} m_{-\bold{k} \pm \bold{k}_s} - m_{\bold{k}}^\dagger m_{-\bold{k} \pm \bold{k}_s}^\dagger \right],
\end{align}
 where $g_{\bold{k}}$ is the magnon-magnon coupling strength such that $g_{\bold{k_q}}=\frac{nL}{s} \frac{\mathcal{B}_{xy} |\mathcal{E}_{xy}|}{2i}$ and we have used $f(t) = \cos(\omega_s t)$. Note that for typical experimental parameters, $\bold{k}_q \gg \bold{k}_s$. To offer practical estimates: for $\omega_q = 2.433$ GHz, the wavenumber is approximately of the order $k_q \approx 10^{7}$ m$^{-1}$. On the other hand, for a phonon velocity of $v_{ph} \approx 6000$ m/s, the corresponding wavenumber is $k_s \approx 10^5$ m$^{-1}$. Therefore, the contributions from $\pm$ terms in Eq.~(\ref{B5}) are approximately identical. Thus, while each of the modes having momentum $\bold{k}$ couples with $\bold{-k}\pm\bold{k}_s$ with different coupling strengths, i.e., $g_{\bold{k}}$ in Eq.~(\ref{B5}), we may approximate these coupling strengths as $g_{\bold{k_q}}$, as in Eq.~(\ref{eq10}), since $\delta (\omega_{\bold{k}}-\omega_{\bold{k_q}})$ emerging from Eqs. (\ref{eq4})-(\ref{eq4.1}) will ensure that only $g_{\bold{k}}=g_{\bold{k_q}}$ contributes to the integrals. We then rewrite Eq.~(\ref{B5}) as
\begin{align}\label{B6}
H_{\text{int}} =  \sum_{\bold{k}} g_{\bold{k}}
\left[ \tilde{m}_{\bold{k}} \tilde{m}_{-\bold{k} + \bold{k}_s} - \tilde{m}_{\bold{k}}^\dagger \tilde{m}_{-\bold{k} + \bold{k}_s}^\dagger \right],
\end{align}
where $\tilde{m}_{\bold{k}} = m_{\bold{k}} / \sqrt{A}$, $A$ is the two-dimensional area with $\left[ \tilde{m}_{\bold{k}}, \tilde{m}_{\bold{k}'}^\dagger \right] = \delta_{\bold{k}, \bold{k}'}$ and we used \[
\sum_{\bold{k}} = \frac{A}{(2\pi)^2} \int d^2\bold{k}.
\] 
Under the evolution of $H_{\text{int}}$, an initial vacuum state evolves into a squeezed vacuum state, i.e.,
\begin{align}\label{B7}
\rho_B = \prod_{\bold{k}} S_{\bold{k}}(\xi) |0_{\bold{k}}\rangle\langle 0_{\bold{k}}|S_{\bold{k}}^\dagger(\xi),
\end{align}
where $S_{\bold{k}}(\xi_k) = \exp \left( \xi_k^* \tilde{m}_{\bold{k}} \tilde{m}_{-\bold{k} + \bold{k}_s} - \xi_k \tilde{m}_{\bold{k}}^\dagger \tilde{m}_{-\bold{k} + \bold{k}_s}^\dagger \right)$ \textcolor{magenta}{}
and $\xi_k = r_{k} e^{i\phi}$. The correlations between any two magnon operators are given by $\langle O_{i}(t)O_j(t^\prime)\rangle=\text{Tr}\left[\rho_BO_i(t)O_j(t^\prime)\right]$ where $O_i$, $O_j$ $\in$ $\{\tilde{m}_{\bold{k}},\tilde{m}_{\bold{k}}^\dagger\}$ $\forall$ $\bold{k}$. Substituting Eq. (\ref{B7}) for $\rho_B$, we obtain the magnon correlations
\begin{align} 
    &\langle \tilde{m}_{\bold{k}}(t)\tilde{m}_{\bold{k}^\prime}(t^\prime)\rangle=M_k e^{-i\omega_{k}(t+t^\prime)}\;\delta^2_{{\bold{k},-\bold{k}^\prime+\bold{k}_s}},\label{B8}
    \\
    &\langle \tilde{m}_{\bold{k}}^\dagger(t)\tilde{m}_{\bold{k}^\prime}^\dagger(t^\prime)\rangle=M_k^* e^{i\omega_k(t+t^\prime)}\;\delta^2_{{\bold{k},-\bold{k}^\prime+\bold{k}_s}},\label{B9}
    \\
    &\langle \tilde{m}_{\bold{k}}^\dagger(t)\tilde{m}_{\bold{k}^\prime}(t^\prime)\rangle=N_ke^{i\omega_k(t-t^\prime)}\;\delta^2_{{\bold{k},\bold{k}^\prime}},\label{B10}
    \\
    &\langle \tilde{m}_{\bold{k}}(t)\tilde{m}_{\bold{k}^\prime}^\dagger(t^\prime)\rangle=(N_k+1)e^{-i\omega_k(t-t^\prime)}\;\delta^2_{{\bold{k},\bold{k}^\prime}},\label{B11}
\end{align}
where $\delta^2_{{\bold{k},\bold{k}^\prime}}$ is the Kronecker delta function. Here, we have assumed that the magnetic system is spatially isotropic, leading to $\omega_{\bold{k}}\approx\omega_{-\bold{k}\pm\bold{k}_s}=\omega_k$. 

The correlations~(\ref{B8})-(\ref{B11}) in the continuum limit lead to Eqs. (\ref{eq11})-(\ref{eq11.3}).  In taking the continuum limit, we replaced $\delta^2({\bold{k}+\bold{k}^\prime}-\bold{k}_s)$ by $\delta^2({\bold{k}+\bold{k}^\prime})$ since the Dirac delta functions will produce the coefficients of the form $a_{\bold{k}}a_{-\bold{k}+\bold{k}_s}\approx a_{\bold{k}}a_{-\bold{k}}$ due to the smallness of $\bold{k}_s$ compared with $\bold{k}=\bold{k}_q$. Therefore, using the correlation relations of the magnon operators in the continuum, it is straightforward to show that
\begin{widetext}
\begin{align}
    \langle B^-_\alpha(t)B^+_\beta(t')\rangle &= \pi \gamma^2 s \int_0^\infty dk \, k^3 e^{-2kd} J_0(k\rho_{\alpha\beta})
    \left(1 + \frac{\sin^2\theta}{2}\right) 
    \Big[N_{k} e^{i\omega_{k}(t-t')} + (N_{k}+1)e^{-i\omega_{k}(t-t')}\Big], \label{B12}  \\
    \langle B^+_\alpha(t)B^-_\beta(t')\rangle &= \pi \gamma^2 s \int_0^\infty dk \, k^3 e^{-2kd} J_0(k\rho_{\alpha\beta}) 
    \left(1 + \frac{\sin^2\theta}{2}\right) 
    \Big[(N_{k}+1)e^{-i\omega_{k}(t-t')} + N_{k} e^{i\omega_{k}(t-t')}\Big],\label{B13}  \\
    \langle B^-_\alpha(t)B^-_\beta(t')\rangle &= \frac{\gamma^2 s}{2} \int d^2{\bold{k}} \, k^2 e^{-2kd} 
    \Big[\cos^2(\phi_{k}) - \sin^2(\phi_{k})\cos^2\theta - i\sin(2\phi_{k})\cos\theta - \sin^2\theta\Big]\nonumber  \\
    &\quad \times e^{i2\phi_{k}} M_{k} e^{i\omega_{k}(t+t')}, \label{B14}  \\
    \langle B^+_\alpha(t)B^+_\beta(t')\rangle &= \frac{\gamma^2 s}{2} \int d^2{\bold{k}} \, k^2 e^{-2kd} 
    \Big[\cos^2(\phi_{k}) - \sin^2(\phi_{k})\cos^2\theta + i\sin(2\phi_{k})\cos\theta - \sin^2\theta\Big] \nonumber \\
    &\quad \times e^{-i2\phi_{k}} M_{k}^* e^{-i\omega_{k}(t+t')}, \label{B15}
\end{align}
\end{widetext}
where $\rho_{\alpha\beta} = |\boldsymbol{\rho_\alpha} - \boldsymbol{\rho_\beta}|$. Notice that in the third and fourth lines of the above correlations, we have only kept the dominant contributions. In particular, when $\theta = 0$, the above equations reduce to
\begin{widetext}
\begin{align}
    \langle B^-_\alpha(t)B^+_\beta(t')\rangle &= \pi \gamma^2 s \int_0^\infty dk \, k^3 e^{-2kd} J_0(k\rho_{\alpha\beta}) 
    \Big[N_{k} e^{i\omega_{k}(t-t')} + (N_{k}+1)e^{-i\omega_{k}(t-t')}\Big], \label{B16}  \\
    \langle B^+_\alpha(t)B^-_\beta(t')\rangle &= \pi \gamma^2 s \int_0^\infty dk \, k^3 e^{-2kd} J_0(k\rho_{\alpha\beta}) 
    \Big[(N_{k}+1)e^{-i\omega_{k}(t-t')} + N_{k} e^{i\omega_{k}(t-t')}\Big], \label{B17} \\
    \langle B^-_\alpha(t)B^-_\beta(t')\rangle &= \pi \gamma^2 s \int_0^\infty dk \, k^3 e^{-2kd} J_0(k\rho_{\alpha\beta}) 
    M_{k} e^{i\omega_{k}(t+t')}, \label{B18} \\
    \langle B^+_\alpha(t)B^+_\beta(t')\rangle &= \pi \gamma^2 s \int_0^\infty dk \, k^3 e^{-2kd} J_0(k\rho_{\alpha\beta}) 
    M_{k}^* e^{-i\omega_{k}(t+t')}.\label{B19}
\end{align}
\end{widetext}
The correlations~(\ref{B16})-(\ref{B19}) together with Eqs.~(\ref{eq4})-(\ref{eq4.1}) are used to calculate the coupling strengths in Eqs. (\ref{eq13})-(\ref{eq17}).

\bibliography{references}
\end{document}